\documentclass{LMCS}
\usepackage{rcp}

\newtheorem*{definition*}{Definition}
\newtheorem*{rule*}{Rule}
\newtheorem*{lemma*}{Lemma}
\newtheorem*{theorem*}{Theorem}

\def\eDef{\pushright{$\blacklozenge$}
    \penalty-700 \par\addvspace{\medskipamount}}
\def\eRul{\pushright{$\blacksquare$}
    \penalty-700 \par\addvspace{\medskipamount}}

\newcommand{\Refines}{\ensuremath{\sqsubseteq}}
\renewcommand{\leadsto}{\ensuremath{\rightsquigarrow}}

\newcommand{\sdef}{~\widehat{=}~}
\newcommand{\iif}{\textbf{if~}}
\newcommand{\oor}{[\!\;\!]}
\newcommand{\Fi}{{\bf fi}}
\newcommand{\Do}{{\bf do~}}
\newcommand{\Od}{{\bf od}}
\newcommand{\imp}{\Rightarrow}
\newcommand{\all}{\forall}
\newcommand{\asgn}{:\!=}
\newcommand{\pl}{\parallel}
\newcommand{\aang}[1]{\langle#1\rangle}
\newcommand{\Angle}[1]{\langle~ #1~ \rangle}
\newcommand{\un}{{\bf ~un~}}
\newcommand{\then}{\rightarrow}
\newcommand{\htr}[3]{\{#1\}~#2~\{#3\}}
\newcommand{\D}{\displaystyle}
\newcommand{\follows}{\Leftarrow}

\newcommand{\code}[5]{
\begin{center}
\begin{tabular}[t]{|l|l|}
\hline
\multicolumn{2}{|l|}{
  \begin{tabular}[t]{@{}l}
    #2
  \end{tabular}
  } \\
\hline 
\begin{minipage}[t]{#1\columnwidth}
#3
\end{minipage}
& 
\begin{minipage}[t]{#1\columnwidth}
#4
\end{minipage}\\ 
\hline
\multicolumn{2}{l}{
  \begin{tabular}[t]{@{}l}
    #5
  \end{tabular}
  } \\
\end{tabular}
\end{center}}

\newcommand{\cccode}[4]{
\begin{center}
\begin{tabular}[t]{|l|l|}
\hline
\multicolumn{2}{|l|}{
  \begin{tabular}[t]{@{}l}
    #2
  \end{tabular}
  } \\
\hline 
\begin{minipage}[t]{#1\columnwidth}
#3
\end{minipage}
& 
\begin{minipage}[t]{#1\columnwidth}
#4
\end{minipage}\\ 
\hline
\end{tabular}
\end{center}}

\def\doi{2 (1:6) 2006}
\lmcsheading%
{\doi}
{25}
{}
{}
{Jan.~12, 2005}
{Mar.~10, 2006}
{}
\begin{document}

\title{Extending the theory of Owicki and Gries with a logic of progress}
\author[B.~Dongol]{Brijesh Dongol\rsuper a}
\address{{\lsuper a}School of ITEE, University of Queensland}
\email{brijesh@itee.uq.edu.au}

\author[D.~Goldson]{Doug Goldson\rsuper b} 
\address{{\lsuper b}\phantom{x}\vskip-\baselineskip}
\email{dgold20@eq.edu.au}

\begin{abstract}
This paper describes a logic of progress for concurrent programs.
The logic is based on that of UNITY, molded to fit a sequential
programming model. Integration of the two is achieved by using
auxiliary variables in a systematic way that incorporates program
counters into the program text. The rules for progress in UNITY
are then modified to suit this new system. This modification is
however subtle enough to allow the theory of Owicki and Gries to
be used without change.
\end{abstract}

\maketitle

\section{Introduction}\label{Intro}
While verifying concurrent programs has been the topic of much
research, deriving them has not. Even less work has been put into
deriving concurrent programs in a way that gives equal
consideration to both progress and safety requirements (as opposed
to derivation that is based only on safety requirements). This
paper contributes to this goal by defining a new logic of safety
and progress. The paper does not address methodological questions
of how to incorporate the logic into a design method for
concurrent program derivation, and this is left as a subject for
further work. The paper confines itself to defining the new logic,
presenting an example of its use, and describing how the logic
compares to other work in this area.

The point of departure for this paper is the theory of Owicki and
Gries \cite{Owicki76,Dijkstra82,Feijen99}, a theory of partial
correctness only, which means that it can only be used to reason
about safety requirements. Two reasons recommend this point of
departure. The first is that this theory is attractively simple.
Proofs are carried out in a programming language (using the
assertional style of Hoare) rather than in some other programming
model such as a Petri net, IO automaton, or process algebra. We
see this as an important advantage for program design, where the
practicality of model-based reasoning turns, in some large part,
on the transparency, ease and reliability of the translation of
the model into code.

The second reason for using the theory is that it has already been
used as an effective method of concurrent program derivation
\cite{Feijen99}, albeit derivation that is based only on safety
requirements. The attitude of Feijen and van Gasteren is
instructive in this regard, as it represents a deliberate decision
to eschew the expressiveness of temporal logic in favour of the
simplicity of Owicki and Gries. The benefit of doing so is a
collection of design heuristics that are attractively simple to
use and that, as already remarked, have been shown to be
effective. The cost of the decision is that reasoning about
progress requirements becomes both informal and post hoc. It is a
welcome outcome that so much can be achieved in this way, yet it
remains true that satisfaction of progress
requirements using this approach is in an important sense left
to chance. The pragmatic attitude of Feijen and van Gasteren,
together with the limitation of the theory of Owicki and Gries,
sets the methodological agenda for this paper. That is, the paper
describes how to extend the theory of Owicki and Gries with a
logic of progress that so far as possible, retains the
simplicity of the original theory while at the same time
provides a logic in which to formalise and prove progress
requirements. This work then is a prolegomenon to our larger goal,
which is a method of program derivation that assigns equal
consideration to both progress and safety requirements.

The step from standard predicate logic to temporal predicate logic
represents an order of magnitude increase in complexity, which is why
Feijen and van Gasteren refused to take it. In their words, ``powerful
formalisms for dealing with progress are available.  However, the
thing that has discouraged $us$ from using them in practice is that
they bring about so much formal complexity. ...  We have decided to
investigate how far we can get in designing multiprograms without
doing $formal$ justice to progress'' (\cite{Feijen99} p79).  Other
authors, while taking the step, fully recognise its significance. For
instance, Lamport writes ``TLA differs from other temporal logics
because it is based on the principle that temporal logic is a
necessary evil that should be avoided as much as possible. Temporal
formulas tend to be harder to understand than formulas of ordinary
first-order logic, and temporal logic reasoning is more complicated
than ordinary mathematical reasoning'' (\cite{Lamport94}, p917).
Caution in the face of this added complexity has recommended to us the
approach taken in UNITY \cite{Chandy88}, where assertion `$P$ leads to
$Q$' formalises an important class of progress requirements called
`eventuality' requirements, and where eventuality assertions are
defined without using temporal logic. The progress logic of UNITY is
ideal for three reasons: the rules fully capture the temporal notion
of leads-to \cite{Gerth89}, they thereby support reasoning about
progress without resort to operational reasoning, and the rules are
simple to use (relative to comparable program logics such as
\cite{Schneider97,Lamport94}).  At the same time, we resile from the
UNITY programming model because it lacks all notion of a control
state, which makes (what should be simple) conventional sequential
programming much harder. Fundamental operators such as ``;'' cannot
easily be represented \cite{Stomp94}.

So we have chosen to add the complexity of the logic of UNITY to
the theory of Owicki and Gries over the complexity of full
temporal predicate logic, or, to be more precise, to add a logic
of progress that, while clearly inspired by UNITY logic, is
tailored to fit the fundamentally different programming model of
multiple sequential programs. In adapting the UNITY logic to fit a
sequential programming model, a decision on how to represent the
control state of a sequential program was first to  be made.
\cite{Owicki76} offers a partial representation of control through
the use of auxiliary variables, while \cite{Schneider97,Lamport87}
opt for a fuller representation through the use of control
predicates. Our approach is a novel use of auxiliary variables to
represent program counters, which provides a complete
representation of the control points in a sequential program. This
means that the formal complications that are introduced by the use
of control predicates in the generalised Hoare logic of
\cite{Schneider97,Lamport87} are avoided in our logic, and we are
able to retain the predicate transformer semantics of Dijkstra.
The main contribution of this paper is to combine the strengths of
these two different theories, Owicki-Gries and UNITY, in order to
create something new.

The paper is structured as follows. Section \ref{OG} describes the
theory of Owicki and Gries and provides background to Section
\ref{ExtOG} which gives the formal basis for the extended logic
described in Section \ref{ProOG}. An application of the new logic
to a program design task is also given in Section \ref{ProOG} and
finally Section \ref{Conc} makes a conclusion.

\section{The theory of Owicki and Gries}\label{OG}

This section describes the theory of Owicki and Gries
\cite{Owicki76,Dijkstra82,Feijen99} as presented in \cite{Feijen99}.
Section \ref{Model} describes the underlying programming language and
its operational model.  Section \ref{CoreOG} describes the predicate
transformer $wlp$ that underlies the logical model of programs and
concludes with the core theory of Owicki and Gries.

\subsection{The programming language and its operational model}\label{Model}

The programming notation is the language of guarded commands
\cite{Dijkstra76}.

\begin{definition*}[Statement]
For statements $S, S_1,
S_2, \dots S_n$, booleans $B_1, B_2, \dots, B_n$, variables $x_1 \dots x_m$ and expressions $E_1 \dots E_m$, a $statement$ is defined inductively as follows.
\begin{enumerate}
\item $skip$ is a statement.
\item A (multiple) assignment $\overline{x \asgn E}$ is a statement
  where, 
  
  $\overline{x \asgn E} \sdef x_1 \asgn E_1 \pl x_2 \asgn E_2 \pl
  \dots \pl x_m \asgn E_m$ \\
  and $x_i \neq x_j$ for $i \neq j$.
\item $S_1;S_2$ is a statement.
\item $\aang{S}$ is a statement.
\item The following are statements, where each $B_i  \then S_i$  is called a {\it guarded
    command} with $guard$ $B_i$ and $command$ $S_i$.
  \begin{enumerate}
  \item $\iif B_1 \then S_1 \oor B_2 \then S_2 \oor \dots \oor B_n \then
    S_n~\Fi$
  \item $\Do B_1 \then S_1 \oor B_2 \then S_2 \oor \dots \oor B_n\then S_n~ \Od$%
  \hbox to182.2 pt{\hfill$\blacklozenge$}
 \end{enumerate}
\end{enumerate}  
\end{definition*}

The statements $IF$ and $DO$ are defined as representatives of the
general notion of an \iif or \Do statement.
\[
\begin{array}[t]{r@{}c@{}l}
IF & \sdef & \iif B_1 \then S_1 \oor B_2 \then S_2~\Fi\\
DO & \sdef & \Do B \then S~ \Od
\end{array}
\]
A {\it sequential program}, also called a {\it component}, is just a
statement. A {\it concurrent program}, also called a {\it
  multiprogram}, is a collection of components, together with a
precondition that defines its initial states. In this paper, we will
refer to a concurrent program as a program and to a sequential program
as a component. The values of the variables in a program define its
current data state. A variable of a component may be $local$ to that
component, meaning it is not read or written by any other component;
$private$ to that component, meaning it can be read but not written by
any other component; or $shared$, meaning it can be both read and
written by any other component.

A component is executed by executing its atomic actions. An atomic
action is an execution step that results in a single update of the
control state of the whole {\it program}, i.e., when an atomic
action is executed, the control state of the component in which
the action occurs changes once, and the control state of all other
components remains the same. Note that an atomic action is
guaranteed to terminate when it is executed. We adopt a
programming model in which an atomic action corresponds to an
assignment statement, to a $skip$ statement, to a guard evaluation
step in an \iif or \Do statement, or to a coarse-grained atomic
statement. The latter is defined by applying the `atomicity
operator' $\aang{S}$ to an arbitrary statement $S$, where the
operator eliminates any control points in $S$ so that $\aang{S}$
is executed atomically as just described. Note that execution of
$\aang{S}$ is only enabled (not blocked) if execution of $S$ is
guaranteed to terminate. While this creates an impossible
difficulty for the implementor, since a machine can not, in
general, decide whether a statement will terminate, the use of
coarse-grained atomic statements in our language allows us to
nicely capture otherwise informal concepts (see \cite{Goldson05}).
\cite{Apt91} solve this problem syntactically, by disallowing $S$
to contain a loop or a synchronisation statement, whereas our
approach is to make it the responsibility of the programmer to
ensure that a coarse-grained atomic statement is guaranteed to
terminate.

Condition synchronisation in the model is achieved using the \iif
statement. Execution of the guard evaluation action of an \iif
statement is blocked when the guard evaluation action is not enabled,
which is when all of the guards are evaluated false. A guard
evaluation action of an \iif statement is therefore a conditional
atomic action because it may not always be enabled. A guard evaluation
action of a \Do statement is an unconditional atomic action because it
is always enabled, as are $skip$ and assignment actions. The
programming model prescribes weak fairness, so that on termination of
an atomic action, an atomic action that follows it, if there is one,
is eventually executed if it is continually enabled. This means that
in the concurrent execution of a number of components, the execution
of the next (continually enabled) atomic action of no component is
delayed indefinitely.

\subsection{Hoare triples, the $wlp$ and
            the core theory of Owicki and Gries}\label{CoreOG}

If $P$ and $Q$ are any two predicates, and $S$ is a statement, a
{\it Hoare-triple},  $\htr{P}{S}{Q}$ is true iff each terminating
execution of $S$ that starts in an initial state satisfying $P$ is
guaranteed to end in a final state satisfying $Q$. $P$ is called
the {\it precondition} of $S$ and $Q$ the $postcondition$.  A
predicate that appears in a Hoare-triple is also called an
$assertion$ and programs that have such assertions are referred to
as being $annotated$. The annotation of a program also defines the
program's initial state with a precondition, which is referred to
as $Pre$.

\begin{definition*}[Weakest Liberal Precondition] The {\it
weakest liberal precondition} ($wlp$) \cite{Dijkstra76} predicate
transformer is defined inductively as follows, where
$P[\overline{x \asgn E}]$ denotes the textual substitution of each
$E_i$ for free occurrences of $x_i$ in $P$.
\begin{enumerate}
\item $wlp.skip.P \equiv P$
\item $wlp.(\overline{x\asgn E}).P \equiv P[\overline{x \asgn E}]$
\item $wlp.\aang{S}.P \equiv wlp.S.P$
\item $wlp.(S_1;S_2).P \equiv wlp.S_1.(wlp.S_2.P)$
\item $wlp.IF.P \equiv (B_1 \imp wlp.S_1.P) \land (B_2 \imp wlp.S_2.P)$
\item $wlp$ in the case of statement $DO$ need no longer be first
  order definable \cite{Gumm99}, as we do not know when (or if) the
  loop terminates. The $wlp$ of a $DO$ loop is defined in terms of a
  countable sequence of conditionals of the form

  \noindent
  $D \sdef \iif B \then S \oor \neg B \then skip~ \Fi$ which gives us: 
  \[wlp.DO.P \equiv \D\bigvee_{n=1}^\infty wlp.D_n.P\] \\
  where $D_n$ is the $n$-fold iteration of statement $D$.\eDef
\end{enumerate}
\end{definition*}

The fundamental relation between Hoare-triples and $wlp$ is that,
for any statement $S$ and predicates $P$ and $Q$\footnotemark,
        \[\htr{P}{S}{Q} \equiv P \imp wlp.S.Q\]

\footnotetext{It is common to relate Hoare-triples to the total
correctness predicate transformer $wp$, however, this is
ill-suited to a programming paradigm in which termination is not
always desired.}

In a program design setting it is usually most convenient to
present proofs using the predicate transformer $wlp$. However,
this is not always the case due to the awkwardness of the
definition of $wlp$ for statement $DO$, where it is more
convenient to make use of the following theorem
\[\htr{P}{DO}{Q} ~~\follows~~ ((P \land B \imp wlp.S.P) \land
                           (P \land \neg B \imp Q))\]

Any predicate $P$ that satisfies this relation is referred to as a
{\it loop invariant}, and proving correctness of an annotated $DO$
statement amounts to finding a $P$ that satisfies this relation.

We are now in a position to state the core theory of Owicki and
Gries, which defines the conditions under which a program
annotation is correct.

\begin{rule*}[Local Correctness] An assertion $P$ in
  a component is {\it locally correct} (LC) when,
\begin{enumerate}
\item if $P$ is textually preceded by program precondition $Pre$, then
  $Pre \imp P$  
\item if $P$ is textually preceded by $\{Q\}~S$, then $\htr{Q}{S}{P}$ holds,
  i.e.,  $Q \imp wlp.S.P$.
\end{enumerate}
\eRul
\end{rule*}

\begin{rule*}[Global Correctness]
An assertion $P$ in a component is {\it globally correct} (GC) if
for each $\{Q\}~S$ from a different component, $\htr{P \land
Q}{S}{P}$ holds, i.e., $P \land Q \imp wlp.S.P$.
\eRul
\end{rule*}

An assertion is {\em correct} if it is both locally and globally correct. An annotation is {\em correct} if all assertions are correct.

\begin{rule*}[Postcondition] A predicate $P$ is a valid postcondition of a
  program if the conjunction of the correct postconditions of the
  components implies $P$.
  \eRul
\end{rule*}

It is useful at this point to introduce a simple example of how
the theory can be used to prove a safety requirement which will
serve to make the foregoing discussion concrete. Consider this
program of two components $X$ and $Y$

\code{.4}
{Program~ (1) \\
$Pre: x=0$ \\
}
{
Component $X$:~~    \\
$
\begin{array}[t]{l}
x \asgn x+1
\end{array}
$
}
{
Component $Y$:~~    \\
$
\begin{array}[t]{l}
x \asgn x+2
\end{array}
$
}
{
$Safety:~$ Program (1) has terminated $\imp x=3$
}

\noindent
Program (1) satisfies $Safety$.

\proof

$X$ and $Y$ are annotated locally correct (LC) and note that both
satisfy part (1) of the LC rule

\cccode{.4}
{Program~ (1) \\
$Pre: x=0$
}
{
Component $X$:~~    \\
$
\begin{array}[t]{l}
\{x=0\}~ \\
~~~x \asgn x+1 \\
\{x=1\}  \\
\end{array}
$
}
{
Component $Y$:~~    \\
$
\begin{array}[t]{l}
\{x=0\}~ \\
~~~x \asgn x+2 \\
\{x=2\}  \\
\end{array}
$
}

\noindent
Global correctness (GC) of the annotation is now arranged by
weakening all four assertions, noting that this maintains LC.

\cccode{.4}
{Program~ (1) \\
$Pre: x=0$
}
{
Component $X$:~~    \\
$
\begin{array}[t]{l}
P:\{x=0 \lor x=2\}~ \\
~~~x \asgn x+1 \\
Q:\{x=1 \lor x=3\}  \\
\end{array}
$
}
{
Component $Y$:~~    \\
$
\begin{array}[t]{l}
\{x=0 \lor x=1\}~ \\
~~~x \asgn x+2 \\
\{x=2 \lor x=3\}  \\
\end{array}
$
}

\noindent
The GC of the assertions $P$ and $Q$ in $X$ are calculated: 

\begin{minipage}[t]{.475\columnwidth}
\begin{derivation}
  \step{}
  \step{wlp.(x \asgn x+2).P}
  \trans{\equiv} {Substituting the value of $P$}
  \step{wlp.(x \asgn x+2).(x=0 \lor x=2)}
  \trans{\follows} {By definition of $wlp$}
  \step{x=0}
  \trans{\equiv} {By logic}
  \step{(x=0 \lor x=2) \land (x=0 \lor x=1)}
  \step{}
\end{derivation}
\end{minipage}%
\begin{minipage}[t]{.475\columnwidth}
\begin{derivation}
  \step{}
  \step{wlp.(x \asgn x+2).Q}
  \trans{\equiv} {Substituting the value of $Q$}
  \step{wlp.(x \asgn x+2).(x=1 \lor x=3)}
  \trans{\follows} {By definition of $wlp$}
  \step{x=1}
  \trans{\equiv} {By logic}
  \step{(x=1 \lor x=3) \land (x=0 \lor x=1)}
  \step{}
\end{derivation}
\end{minipage}

\noindent
Finally, the conjunction of the two final assertions of $X$ and
$B$ establishes the desired safety requirement

\begin{derivation}
    \step{(x=1 \lor x=3) \land (x=2 \lor x=3)~~ \equiv~~ x=3
    \hbox to224 pt{\hfill\qEd}}
\end{derivation}
\medskip

The simplicity of the core theory is reflected in its limited
power. The lack of a  means to reason about a program's control
state means that safety requirements that are clearly met may not
be provable, such as in the following program.

\code{.4}
{
Program~ (2)\\
$Pre:~ x=0$
}
{ Component X: \\
  $
  \begin{array}[t]{l}
    x \asgn x+1
  \end{array}
  $
}
{ Component Y: \\
  $
  \begin{array}[t]{l}
    x \asgn x+1
  \end{array}
  $
}
{
$Safety:~$ Program (2) has terminated $\imp x=2$
}

It is an interesting exercise to convince yourself that this safety
requirement is not provable in the core theory. The solution in
\cite{Owicki76} is to add auxiliary information into a program which
could be used in its correctness proof. We start by defining an {\em
  auxiliary assignment}, which is an assignment to a fresh variable,
different from all program variables, called an {\em auxiliary
  variable}. The assignment may only appear as part of an atomic
action, hence, does not introduce any new control points. We require
that actions remain well-formed when all auxiliary assignments are
removed.  Furthermore, as addition of auxiliary information should not
affect control and data states of the original program, auxiliary
variables may not appear in any guard and assigned to a non-auxiliary
variable.



Returning to the example of Program (2), we augment the program
with auxiliary assignments to fresh variables $pc.A$ and $pc.B$ to
give us the following program.

\code{.4}
{
Program~ (3)\\
$Pre:~ x=pc.X=pc.Y=0$ 
}
{ Component X: \\
  $
  \begin{array}[t]{l}
    x \asgn x + 1 \pl pc.X \asgn 1
  \end{array}
  $
}
{ Component Y: \\
  $
  \begin{array}[t]{l}
    x \asgn x+1 \pl pc.Y \asgn 1
  \end{array}
  $
}
{
$Safety:~$ Program (3) has terminated $\imp x=2$ 
}

\noindent
Program (3) satisfies $Safety$.

\proof

This is now much as for Program (1). The two components can be
annotated for LC

\cccode{.4}
{
Program~ (3)\\
$Pre:~ x=pc.X=pc.Y=0$ 
}
{ Component X: \\
  $
  \begin{array}[t]{l}
    \{x=0\}\{pc.X=0\}\\
    ~~~x \asgn x + 1 \pl pc.X \asgn 1\\
    \{x=1\}\{pc.X=1\}
  \end{array}
  $
}
{ Component Y: \\
  $
  \begin{array}[t]{l}
    \{x=0\}\{pc.Y=0\}\\
    ~~~x \asgn x+1 \pl pc.Y \asgn 1\\
    \{x=1\}\{pc.Y=1\}
  \end{array}
  $
}

\noindent
GC is arranged by a combination of strengthening and weakening
these assertions as follows: 

\begin{small}
\cccode{.47}
{
Program~ (3)\\
$Pre:~ x=pc.X=pc.Y=0$ 
}
{ Component X: \\
  $
  \begin{array}[t]{@{}l@{~}l}
    P: & \{(x=0 \land pc.Y=0) \lor (x=1 \land pc.Y=1)\} \\
       & \{pc.X=0\} \\

       & \quad x \asgn x + 1 \pl pc.X \asgn 1\\

    Q: & \{(x=1 \land pc.Y=0) \lor (x=2 \land pc.Y=1)\} \\
       & \{pc.X=1\}\\
  \end{array}
  $
}
{ $~~~$Component Y: \\
  $
  \begin{array}[t]{l}
     \{(x=0 \land pc.X=0) \lor (x=1 \land pc.X=1)\}\\\{pc.Y=0\}
     \\

    \quad x \asgn x+1 \pl pc.Y \asgn 1\\
    \{(x=1 \land pc.X=0) \lor (x=2 \land pc.X=1)\}\\
    \{pc.Y=1\}
  \end{array}
  $
}
\end{small}

\noindent
As before, the GC of $P$ and $Q$ in $X$ are calculated:
\begin{derivation}
  \step{wlp.(x \asgn x+1 \pl pc.Y \asgn 1).
    (((x=0 \land pc.Y=0) \lor (x=1 \land pc.Y=1)) \land pc.X=0)}
  \trans{\equiv} {By definition of $wlp$}
  \step{x=0 \land pc.X=0}
  \trans{\follows} {By logic}
  \step{((x=0 \land pc.Y=0) \lor (x=1 \land pc.Y=1)) \land pc.X=0
    \land pc.Y=0} 
\end{derivation}
\smallskip
\begin{derivation}
  \step{wlp.(x \asgn x+1 \pl pc.Y \asgn 1).
    (((x=1 \land pc.Y=0) \lor (x=2 \land pc.Y=1)) \land pc.X=1)}
  \trans{\equiv} {By definition of $wlp$}
  \step{x=1 \land pc.X=1}
  \trans{\follows} {By logic}
  \step{((x=1 \land pc.Y=0) \lor (x=2 \land pc.Y=1)) \land pc.X=1
    \land pc.Y=0} 
\end{derivation}
\smallskip
\noindent Finally, the conjunction of the two final assertions of $A$ and
$B$ establishes the desired safety requirement.

\begin{derivation}
    \step{((x=1 \land pc.Y=0) \lor (x=2 \land pc.Y=1)) \land pc.Y=1 ~~\imp ~~ x=2%
    \hbox to120 pt{\hfill\qEd}}
\end{derivation}
\medskip

Noting that Program (3) is just Program (2) with auxiliary
assignments to $pc.A$ and $pc.B$ superimposed on it, we are
entitled to conclude that Program (2) satisfies the same safety
requirement, because the two programs are equivalent in having the
same data and control states.

\section{The extended theory of Owicki and Gries}\label{ExtOG}

It is fairly clear, so far as reasoning about progress is
concerned, that the theory of Owicki and Gries is deficient
because it lacks a systematic means to describe a program's
control state. Any extension to the theory must therefore provide
for this, and the extension to be described in this section has
two parts. First, control points in a component are named by
naming the atomic action to be executed at the corresponding
point. This is done by labelling all of the atomic actions in the
component. Second, an auxiliary variable is introduced into each
component in a way that models its `program counter', i.e., the
value of this variable indicates the active control point in the
component, which is just the label of the atomic action that
corresponds to that control point.

Sections \ref{Labels} and \ref{Counters} introduces
the twin notions of a labelled program and a program counter while
Section \ref{Control} reviews the reasons why program counters
were chosen over control predicates as the formalisation of
program control states.

\subsection{Labelled statements}\label{Labels}

The first step toward describing an active control point in a
statement requires being able to refer to the next atomic action
to be executed. We do this by assigning a unique label to
each atomic action that occurs in the statement. The label of a
statement's initial atomic action will be called the {\it initial
label} of that statement. In addition, a label will be assigned to
the end of the statement which will be called the {\it final
label} of the statement. A final label of a statement will always
label the initial atomic action of a statement that follows it.
However, if there is no following statement, then the final label
does not refer to any atomic action, but simply marks the end of
the statement.

\begin{definition*}[Labelled Statement]\

\begin{enumerate}
\item A labelled $skip$ statement has the form $i: skip~ j:$
where $i$ is the initial label and $j$ is the final label.

\item A labelled assignment statement has the form
$i: x\asgn E~ j:$ where $i$ is the initial label and $j$ is the
final label.

\item A labelled sequential statement has the form $i: S_1; j: S_2~ k:$
where $i$ is the initial label of statement $S_1$, $j$ is the
final label of $S_1$, $j$ is the initial label of statement $S_2$
and $k$ is the final label of $S_2$.

\item A labelled coarse-grained atomic statement $\aang {S}$ has the form
$i: \aang {S}~ j:$ where $i$ is the initial label and $j$ is the
final label, and statement $S$ is not labelled.

\item A labelled statement $IF$ has the form

        $ \quad i: \iif B_1 \then~ j: S_1 \oor
                   B_2 \then~ k: S_2 ~\Fi~ l: $\\
where $i$ is the initial label of $IF$ and $l$ is the final label
of $IF$, $i$ is the label of the initial atomic action of $IF$,
which is the guard evaluation action, and $j$ and $k$ are the
final labels of the guard evaluation action. $j$ is the initial
label of statement $S_1$, $k$ is the initial label of statement
$S_2$ and $l$ is the final label of both $S_1$ and $S_2$.

\item A labelled statement $DO$ has the form

        $\quad i: \Do B \then~ j: S~ \Od~ k:$\\
where $i$ is the initial label of $DO$ and $k$ is the final label
of $DO$, $i$ is the label of the initial atomic action of $DO$,
which is the guard evaluation action, and $j$ and $k$ are final
labels of the guard evaluation action. $j$ is the initial label of
statement $S$ and $i$ is the final label of $S$.

\item If $i$ and $j$ are the initial labels for 
different actions of any statement, then $i \neq j$.\eDef
\end{enumerate}
\end{definition*}

In what follows $A.i$ will be used to denote `the atomic action in
component $A$ labelled $i$' whenever $i$ is not the final label of
component $A$.

\subsection{Modelling program counters}\label{Counters}

There are essentially two ways of using this additional information
that labelled statements provide. One way is to introduce into the
logic new control predicates to express propositions such as, for
instance, that `control in component $A$ is at the atomic action
labelled $i$'. This kind of approach is taken in (\cite{Schneider97},
pp96-108, pp136-140) and in (\cite{Lamport87}), but the cost is that
the familiar axioms of Hoare logic, as presented in Section
\ref{CoreOG}, must be given up in favour of generalised axioms that
take account of the fact that, say, $\htr{P}{i: skip~j:}{P}$ is no
longer true for all $P$ (for example, when $P$ asserts that `control
is at label $i$').  A further cost is that new axioms must be
introduced to capture the intended interpretation of the new control
predicates. The desire to make only conservative extension to the
theory of Owicki and Gries, prompted by the desire to retain old,
familiar and trusted ways (the $wlp$), has led us to resist this
approach in favour of the use of auxiliary variables to reason about
the control state. In our method, $\htr{P}{i:skip~j:}{P}$ is also not
an axiom. However, we do avoid the extra axioms on the control state
required by control predicates, and are able to retain the $wlp$ as
the main tool for proving predicates.

Consequently, we formalise a program's control state in the
following way. An auxiliary variable is introduced into each
component in a way that models its `program counter', i.e., the
value of this variable indicates the active control point in the
component, which is just the label of the next atomic action to be
executed, or the end of the component if no such action exists.
Given a component $A$, this variable $pc.A$ will be called the
{\it program counter} of $A$, and its essence is to record the
control state of $A$, but in so doing to change neither the
program's control state nor its data state. Given this essence,
program counter $pc.A$ must be updated at every atomic action in
$A$ in a way that assigns $pc.A$ a final label of that action.
This is done by superimposing an auxiliary assignment to $pc.A$
onto every atomic action in $A$ in the following way.

\begin{definition*} [Program Counter] Given a  program with
precondition $Pre$ and labelled component $A$, variable $pc.A$ is
the {\it program counter} of $A$ when

\begin{enumerate}
\item $pc.A$ is a local variable of $A$.

\item If $i$ is the initial label of $A$ then $Pre \imp pc.A=i$

\item A labelled $skip$ statement has the form
$i: \aang{skip;pc.A\asgn j}~ j:~$.

\item A labelled assignment statement has the form
$i: x\asgn E \pl pc.A \asgn j~ j:~$.

\item A labelled coarse-grained atomic statement has the form
$i: \aang {S;pc.A\asgn j}~ j:~$.

\item A labelled statement $IF$ has the form

  $\quad i: \iif \aang{B_1 \then~ pc.A\asgn j}~ j: S_1 \oor
  \aang{B_2 \then~ pc.A\asgn k}~ k: S_2 ~\Fi~ l: $

\item A labelled statement $DO$ has the form

  $ \quad i: \Do \aang{B \then~ pc.A\asgn j}~ j: S~
  \oor \aang{\neg B \then  pc.A\asgn k}~ \Od~ k:$
\end{enumerate}
\end{definition*}

Given that guard evaluation is an atomic action (as it changes the
program control state whenever a guard is evaluated $true$), and
given that a program counter must be updated at {\it every} atomic
action in a component, we are required to extend the grammar of
statements $IF$ and $DO$ in order to make explicit the state
change that can accompany a guard evaluation. To this end we
modify the syntax of guarded command $B \then j:S$ to $\aang{B
\then pc.A \asgn j}~ j:S$ so that the transfer of program control
from the guard evaluation to the initial action of $S$ when guard
$B$ evaluates $true$ is made explicit. Note how atomicity brackets $\Angle{}$ are used to indicate that
the update of the program counter is part of the guard evaluation. However, we acknowledge that this grammar is awkward,
because it is semantically misleading whenever a statement
consists of several alternatives. For example, in statement
\[\iif \aang{B_1 \then~ pc.A\asgn j}~ j: S_1 \oor
            \aang{B_2 \then~ pc.A\asgn k}~ k: S_2 ~\Fi~ l: 
\]
\noindent
the two pairs of atomicity brackets suggest two atomic guard
evaluations, which is not the case, rather there is one atomic
guard evaluation labelled by $i$, which has three outcomes, the
first where guard $B_1$ is evaluated to $true$ and control passes
to the initial action of $S_1$ labelled by $j$, the second where
guard $B_2$ is evaluated to $true$ and control passes to the
initial action of $S_2$ labelled by $k$, and the third where both guards 
$B_1$ and $B_2$ are evaluated to
$false$ and control remains at the guard evaluation action
labelled by $i$.

The case of statement $DO$
\[i: \Do B \then~ j: S~ \Od~ k:
\]
\noindent
is further complicated by the fact that the loop is not a blocking
statement, which is to say that when guard $B$ is evaluated
$false$ control does not remain at the guard evaluation labelled by $i$, but rather it passes to the control point
labelled by $k$. This transfer of control requires an explicit
update to the program counter, which we have accommodated by
changing the grammar of the $DO$ statement in a way that makes
this outcome of the guard evaluation explicit
\[
i: \Do B \then~  j: S~ \oor \neg B \then~ \Od~ k:
\]
The $DO$ statement now admits a guarded command $\neg B \then$
with an empty command, which, if selected, has the total effect on
the program state of passing control to the control point labelled
by $k$. The operational semantics of this syntactically modified
$DO$ statement is unchanged, with the sole purpose of the
modification being to introduce a peg on which to hang the
assignment $pc.A \asgn k$.

Finally, note that we are free to interpret predicate $pc.A=i$ to
mean that `control in $A$ is at $A.i$' because $pc.A=i$ is a
correct precondition of $A.i$ and because labels are unique. LC
follows from the definition of $pc.A$, and GC follows from the
same, on account of $pc.A$ being a local variable of $A$.

\subsection{Program counters vs. control predicates}\label{Control}

Recalling that the reason for choosing program counters over
control predicates has been driven by a desire to make only
conservative changes to the theory of Owicki and Gries, we can
view this choice as one of a superficial (i.e., syntactic) change
to guarded commands in order to make explicit the way that a guard
evaluation can change the control state, over a significant
(i.e., semantic) change to the program logic. The chief practical
gains are that we are able to retain the semantics of $wlp$ as the
logical basis of the programming model and that the absence of
primitive control predicates means that we do not need to
introduce additional logical rules to define them. The core theory
of Owicki and Gries as described in Section \ref{CoreOG} therefore
remains the same under the changes described in Sections
\ref{Labels} and \ref{Counters}, and the definition of the $wlp$
predicate transformer is extended to a labelled statement with
program counter $pc$ as follows

\begin{enumerate}
\item $wlp.(i: \aang{skip;pc\asgn j}~ j:).P \equiv wlp.(pc \asgn j).P$
\item $wlp.(i: (\overline{x \asgn E} \pl pc \asgn j)~ j:).P \equiv P[\overline{x \asgn E} \pl pc \asgn j]$
\item $wlp.(i: \aang {S;pc \asgn j}~ j:).P \equiv wlp.(S;pc \asgn j).P$
\item $wlp.(i: S_1; j: \ S_2~ k:).P \equiv wlp.(i: S_1~ j:).(wlp.(j: S_2~ k:).P)$
\item 
  $\begin{array}[t]{@{}l}
    wlp.(i: \iif \aang{B_1 \then~ pc\asgn j}~ j: S_1 \oor
    \aang{B_2 \then~ pc\asgn k}~ k: S_2 ~\Fi~ l:).P
    \\
    \equiv
    \\
    (B_1 \imp wlp.(pc \asgn j).(wlp.(j:S_1~l:).P)) ~\land~
    (B_2 \imp wlp.(pc \asgn k).(wlp.(k:S_2~l:).P))
  \end{array}$
\item $\begin{array}[t]{@{}l}
        \htr{P}{i: \Do \aang{B \then~ pc\asgn j}~ j: S~
          \oor \aang{\neg B \then  pc\asgn k}~ \Od~ k:}{Q}
        \\
        \follows
        \\
        (P \land B \imp wlp.(pc \asgn j).(wlp.(j:S~i:).P)) \land
        (P \land \neg B \imp wlp.(pc \asgn k).Q)
    \end{array}$
\end{enumerate}

It is noteworthy that typical axioms \cite{Lamport87, Alpern89} that
are required to define the meaning of a control predicate now become
easy derived rules of the program counters model.

\begin{enumerate}
\item Each component has at most one active control point.  This is
  trivial on account of \\$(\all i,j~: j \neq i:~ pc.A=i \imp pc.A
  \neq j)$ and the uniqueness of labels.
\item Each component has at least one active control point.  This
  holds on account of the invariance of $(\exists i~:\!:~ pc.A=i)$.
\item Execution of an atomic statement in component a different from
  $A$ does not change the active control point in component $A$. This
  is trivial on account of $A$'s program counter being a local
  variable of $A$.
\end{enumerate}

Against the advantages of using program counters, the chief
drawback is the syntactic complexity that the program counter
assignments add to the program under consideration. However, this
added complexity is nicely avoided in practice by making the
assignments implicit in the program. In effect, this amounts to
redefining the $wlp$ for a labelled statement with implicit
program counter $pc$ as follows

\begin{enumerate}
\item $wlp.(i: skip~ j:).P \equiv wlp.(pc \asgn j).P$
\item $wlp.(i: \overline{x \asgn E}~ j:).P \equiv P[\overline{x \asgn E} \pl pc \asgn j]$
\item $wlp.(i: \aang{S}~ j:).P \equiv wlp.(S;pc \asgn j).P$
\item $wlp.(i: S_1; j: \ S_2 ~ k:).P \equiv wlp.(i: S_1~ j:).(wlp.(j: S_2~ k:).P)$
\item $\begin{array}[t]{@{}l}
    wlp.(i: \iif B_1 \then~ j: S_1 \oor
    B_2 \then~ k: S_2 ~\Fi~ l:).P
    \\\equiv\\
    (B_1 \imp wlp.(pc \asgn j).(wlp.(j:S_1~l:).P)) \land
    (B_2 \imp wlp.(pc \asgn k).(wlp.(k:S_2~l:).P))
  \end{array}$
\item 
  $\begin{array}[t]{@{}l}
    \htr{P}{i: \Do B \then~ j: S~  \Od~ k:}{Q}
    \\ \follows \\
    (P \land B \imp wlp.(pc \asgn j).(wlp.(j:S~i).P)) \land
    (P \land \neg B \imp wlp.(pc \asgn k).Q)
  \end{array}$
\end{enumerate}

and this is what we do.

\section{A logic of progress for the extended theory}\label{ProOG}

As we now have the means to reason about the control state of a
program, we are now in a position to extend the theory to support
reasoning about progress requirements. The rules for progress in
the extended theory are described in Section \ref{ruleprog}.
Section \ref{IP} describes an application of the new logic to a
program design task, which compares favourably to the treatment in
(\cite{Feijen99}, pp2 07-212) and Section \ref{GCL} describes a
second application of the logic, this time to the proof of
correctness of a program transformation.

\subsection{Rules of progress}
\label{ruleprog}

As already remarked in Section \ref{Intro}, the logic to be presented
is almost just that of UNITY (\cite{Chandy88}, pp47-74), where the
notion of progress is formalised using the relation $leads$-$to$
(denoted $\leadsto$), where, for any predicates $P$ and $Q$, $P
\leadsto Q$ holds if it is always the case that in a program state in
which $P$ holds, execution of the program is such that a program state
will eventually be reached in which $Q$ holds. In temporal logic
\cite{MP92} terms, $P \leadsto Q \equiv \Box (P \imp \Diamond Q)$
where $\Box$ and $\Diamond$ are the `always' and `eventually'
operators respectively. In order to axiomatize this relation, we begin
by defining the notion of $unless$ ($\!\!$\un$\!\!$).

\begin{definition*} [Unless]
  If $P$ and $Q$ are any two predicates, $P \un Q$
  holds if
  \[\htr{P \land \neg Q \land U}{S}{P\lor Q} \]
  holds for all atomic statements $\{U\}~S$, where $U$ denotes the
  precondition of $S$ in the annotated program.
  \eDef
\end{definition*}
Relation $unless$ says that a program state in which $P$ holds and $Q$
does not, is perpetuated until a state is reached in which $Q$ holds.
But note that this does not guarantee that $Q$ will ever hold, for (an
extreme) example, $true \un Q$ holds for all $Q$, including $false$.
To formalise progress properties we also need a notion of what it
means for a statement to establish a predicate given that it is not
yet true. In (\cite{Chandy88}, pp50-52) this is formalised by the
relation $ensures$, which forms the basis of the definition of
$leads$-$to$. In our setting, and purely for presentational reasons,
we have chosen not to define $ensures$, but rather to define the basic
part of $leads$-$to$ directly in terms of the several forms of atomic
action in the programming language. More substantially, the basic part
of our definition of $leads$-$to$, which is the point at which the
relation is bound to the program under consideration, is the only
point at which the two definitions of $leads$-$to$ differ, the
inductive part of our definition being identical to that in
(\cite{Chandy88}, p52). However, this difference in the basic
definition of $leads$-$to$ is an essential difference, on account of
the fundamentally different programming model that is used here and in
UNITY. Our programs consist of a number of concurrently executing
sequential components, each of which is constructed using the guarded
command language \cite{Dijkstra76}, whereas a UNITY program is a single
non-terminating loop of guarded assignments.

We remind ourselves that for the basic part of the definition of
$leads$-$to$, the atomic actions are $skip$, assignment, guard
evaluation and coarse-grained atomic statements of the form $\aang{S}$
for arbitrary statement $S$. A judgment $P \leadsto Q$ arrived at
using this rule ensures that if a program state is reached in which
$P$ holds, execution of the program is such that $P$ will continue to
hold until a program state is reached in which $Q$ holds, and,
further, a state in which $Q$ holds will be reached. We will call this
the `immediate progress' rule because it allows us to actually exhibit
an atomic action that is guaranteed to bring $Q$ about. Our convention
is that operator $\leadsto$ binds weaker than any logical operator.
Hence, for example, $(P \imp Q) \leadsto R ~\equiv~ P \imp Q \leadsto
R$.

\begin{rule*}[Immediate Progress] $P \leadsto Q$ holds if
$P\un Q$ holds, and there exists a labelled statement with initial
label $i$ in a component with program counter $pc$ and

\begin{enumerate}
\item $P \land \neg Q \imp pc=i$
\item
    \begin{enumerate}
    \item the statement is a $skip$ or an assignment statement $i: S~ j:$ and,
      \begin{itemize}
      \item [] $P \land \neg Q ~\imp~ wlp.(i: S~ j:).Q$.
      \end{itemize}
    \item the statement is a coarse-grained atomic statement
	$i: \langle S \rangle~j:$ and,
	\begin{itemize}
	\item [] $P \land \neg Q ~\imp~ wp.S.(Q[pc \asgn j])$
	\end{itemize}
    \item the statement is an $IF$ statement $i: \iif B_1 \then~ j:
      S_1 \oor B_2 \then~ k: S_2 ~\Fi~ l:$ and,
      \begin{enumerate}
      \item $P \land \neg Q ~\imp~ B_1 \lor B_2$
      \item $(P \land \neg Q \land B_1 \imp Q[pc \asgn j])) ~~\land~~(P \land \neg Q \land B_2 \imp Q[pc \asgn k])$.
      \end{enumerate}

    \item the statement is a $DO$ statement $i: \Do B \then~ j: S~
      \Od~ k:$ and,

      $(P \land \neg Q \land B \imp Q[pc \asgn j]) ~~\land~~ (P \land
      \neg Q \land \neg B \imp Q[pc \asgn k])$.\hbox to100 pt{\hfill$\blacksquare$}
    \end{enumerate}
\end{enumerate}  
\end{rule*}

To make sense of this rule we provide these interpretative notes.  $P
\leadsto Q$ is here justified on the basis of being able to actually
exhibit a continually enabled atomic action at an active control point
that makes $Q$ true when it is executed. To see how the rule
formalises this, we first note that $P \land \neg Q$ is assumed.  As
$P \un Q$ must hold, we can be assured that $P$ remains true as long
as $\neg Q$ is true. Clause 1 establishes that control is at an atomic
action labelled $i$ in a component. Clause 2 establishes that this
action is enabled when $P \land \neg Q$ is true, and that its
execution makes $Q$ true. It follows from clause 1 that the action is
continually enabled as long as $\neg Q$ is true and as we are assuming
weak fairness, that the action is eventually executed. Clause 2 is
separated into three cases to cover the three kinds of atomic actions:
execution of an atomic statement; guard evaluation to an \iif
statement; and guard evaluation to a \Do statement. In case (2a), an
assignment action is always enabled and it is enough to ensure that
its execution makes $Q$ true. In case (2b), a guard evaluation action
in an \iif statement is not always enabled and so clause (2bi) ensures
that it is enabled when $P \land \neg Q$ is true.  Clause (2bii)
further ensures that its execution makes $Q$ true.  In case (2c), a
guard evaluation action in a \Do statement is always enabled and it is
again enough to ensure that its execution makes $Q$ true.


The inductive part of the definition of $leads$-$to$ is given by

\begin{rule*}[Inductive Progress]\

  \begin{enumerate} 
  \item [] (Transitivity) $(P \leadsto R) ~\follows~ (P \leadsto Q)
    \land (Q \leadsto R)$
  \item [] (Disjunction) For any set $W$, $((\exists i~: i \in W:~ P.i)
    \leadsto Q) ~\follows~ (\all i~: i\in W:~ P.i \leadsto Q)$
  \end{enumerate}
  \eRul
\end{rule*}

The rule of transitivity requires no explanation. The rule of
disjunction, in its finite application of, say, two progress
assertions, amounts to the inference that if $P.0 \leadsto Q$ and $P.1
\leadsto Q$ then $P.0 \lor P.1 \leadsto Q$. Via a finite number of
applications of the immediate and inductive progress rules, we are now
able to prove any eventuality property that can be proved using
leads-to. The `next' temporal operator is missing from our logic, just
as it was missing from UNITY. However, this is not a big problem in
concurrent environments as reasoning about `next' seldom makes sense
to the inherent non-determinacy.

\cite{Chandy88}
also present a thorough treatment of a collection of derived rules
for leads-to, all of which remain true in our setting, and
which are listed below. The proofs of these derived rules are
presented in Appendix A.

\begin{rule*}[Derived Progress Rules]\

  \begin{enumerate}
  \item (Implication Theorem) 
    $P \leadsto Q ~\follows~ (P \imp Q)$ 
  \item (Impossibility Theorem) 
    $\neg P ~\follows~ (P \leadsto false)$ 
  \item (Disjunction Theorem) 

    $((\exists m ~: m \in W :~ P.m) \leadsto (\exists m ~: m \in W:~
    Q.m)) ~\follows~ (\all m~: m \in W :~P.m \leadsto Q.m)$
  \item (Cancellation Theorem) 

    $(P \leadsto Q \lor R) ~\follows~ (P\leadsto Q \lor D) \land (D
    \leadsto R)$
  \item (PSP (Progress-Safety-Progress) Theorem) 

    $(P \land R \leadsto (Q \land R) \lor D) ~\follows ~ (P \leadsto
    Q) \land (R \un D)$
  \item (Induction Theorem) Let $M$ be a total function from program
    states to set $W$. Let $(W,\prec)$ be well-founded. Variable $m$ in
    the following premiss ranges over $W$ and predicates $P$ and $Q$
    do not contain free occurrences of variable $m$. Then, 

    $(P \leadsto Q) ~\follows~ (\all m~:\!:~ P \land M=m \leadsto (P
    \land M\prec m) \lor Q)$
  \item (Completion Theorem) Let $P.i$ and $Q.i$ be predicates where
    $i$ ranges over a finite set. Then, 
    
    $((\all i~:\!:~ P.i) \leadsto (\all i~:\!:~ Q.i) \lor D)
    \follows~ (\all i~:\!:~ (P.i \leadsto Q.i \lor D) \land (Q.i \un
    D)) $\hbox to66.6 pt{\hfill$\blacksquare$}
  \end{enumerate}
\end{rule*}

The remainder of this section gives two examples of how the new
logic can be used. The first presents an application of the logic
to a program design task, which compares favourably to the
treatment in (\cite{Feijen99}, pp207-212), and the second presents
a proof of correctness of a program transformation called the
``guard conjunction lemma'', which is taken from the same source
(\cite{Feijen99}, pp118-120).

\subsection{The initialisation protocol}\label{IP}

The first example is taken from  \cite{Feijen99} where it appears
as both an exercise in verification  (p84) and as an exercise in
design (p207). Here we present an alternative design that starts
with the following program

\code{.4}
{
The Initialisation Protocol \\
$Pre:~ true$ \\
}
{ Component X: \\
  $
  Init.X;~ \\
  y \asgn false;~ \\
  \aang{\iif y \then~ skip~\Fi};~ \\
  S.X
  $
}
{ Component Y: \\
  $
  \begin{array}[t]{l}
    Init.Y;~ \\
    x \asgn false;~ \\
    \aang{\iif x \then~ skip~\Fi};~ \\
    S.Y
  \end{array}
  $
}
{
$Progress:~ $ There is no individual deadlock
}

The safety requirement of the initialisation protocol is omitted
from the specification on account of this program already
satisfying it, the requirement being that $X$ cannot begin
execution of code $S.X$ until $Y$ has completed execution of code
$Init.Y$, and vice versa. This requirement is maintained provided
that only assignment $y \asgn true$ in $Y$ is allowed, and this
only  after $Init.Y$, which is nicely ensured by restricting
attention to the protocol code below:

\code{.4} { The Initialisation Protocol -- simplified and labelled \\
$Pre:~ pc.X=pc.Y=1$ \\
} 
{ Component X: \\
  $
  1:~y \asgn false;~ \\
  2:~ \aang{\iif y \then~ skip~\Fi}~ \\
  3:~
  $
}
{ Component Y: \\
  $
  \begin{array}[t]{l}
    1:~x \asgn false;~ \\
    2:~ \aang{\iif x \then~ skip~\Fi}~ \\
    3:~
  \end{array}
  $
}
{
$Progress: ~pc.X=2 \leadsto pc.X=3$ \\
$Topology: ~$ Only $y\asgn true$ is allowed in $Y$
}

\noindent
$Progress$ is proved as follows:

\begin{derivation}
  \step{pc.X=2}
  \trans{\leadsto}{By case analysis on the guard of $X.2$}
  \step{(pc.X=2 \land y) \lor (pc.X=2 \land \neg y)}
  \trans{\leadsto}{Immediate progress rule with $X.2$, $y$
    is GC in $X$} 
  \step{pc.X=3 \lor (pc.X=2 \land \neg y)}
\end{derivation}

\noindent
As this is our first encounter with the immediate progress rule, let
us elaborate the last step of the proof. We have:
\begin{equation}
  \label{eq:0}
  (pc.X=2 \land y) \lor (pc.X=2 \land
\neg y) ~\leadsto~ pc.X=3 \lor (pc.X=2 \land \neg y)
\end{equation}

\noindent
This means we have the following instantiations: 

\begin{tabular}[t]{l}
  $P \sdef (pc.X=2 \land y) \lor (pc.X=2 \land \neg y)$ \\
  $Q \sdef pc.X=3 \lor (pc.X=2 \land \neg y)$. 
\end{tabular}
\\
\\
We first show that $P \un Q$ holds.\\

\noindent
\begin{minipage}[t]{.485\columnwidth}
Against $X.1$ we have:

\begin{derivation}
  \step{P \land \neg Q \:\imp\: wlp.(X.1).(P \lor Q)}
  \trans{\equiv}{Substituting value of $X.1$ and by $wlp$}
  \step{P \land \neg Q \:\imp\: (P\! \lor\! Q)[y \asgn\! false, pc.X \asgn
    2]}
  \trans{\equiv}{By substitution}
  \step{P \land \neg Q \:\imp\: true}
  \trans{\equiv}{By logic}
  \step{true}
  \step{}
\end{derivation}

Against $Y.1$ we have:

\begin{derivation}
  \step{P \land \neg Q ~\imp~ wlp.(Y.1).(P \lor Q)}
  \trans{\equiv}{Substituting value of $Y.1$ and by $wlp$}
  \step{P \land \neg Q ~\imp~ P \lor Q}
  \trans{\equiv}{By logic}
  \step{true}
  \step{}
\end{derivation}

\end{minipage}
\begin{minipage}[t]{.485\columnwidth}
Against $X.2$ we have:

\begin{derivation}
  \step{P \land \neg Q ~\imp~ wlp.(X.2).(P \lor Q)}
  \trans{\equiv}{Substituting value of $X.2$ and by $wlp$}
  \step{P \land \neg Q \land y ~\imp~ (P \lor Q)[pc.X \asgn 3]}
  \trans{\equiv}{By $wlp$ calculation}
  \step{P \land \neg Q \land y ~\imp~ true}
  \trans{\equiv}{By logic}
  \step{true}
  \step{}
\end{derivation}

Against $Y.2$ we have:

\begin{derivation} 
  \step{P \land \neg Q ~\imp~ wlp.(Y.2).(P \lor Q)}
  \trans{\equiv}{Substituting value of $Y.2$ and by $wlp$}
  \step{P \land \neg Q \land x~\imp~ P \lor Q}
  \trans{\equiv}{By logic}
  \step{true}
  \step{}
\end{derivation}
\end{minipage}

\noindent 
Clause (1) of the immediate progress rule holds, as $P \land \neg Q
\imp pc.X = 2$. Finally, as we are dealing with a coarse-grained
atomic statement, we refer to clause $(2a)$ which gives us:

\begin{derivation}
  \step{P \land \neg Q ~\imp~ wlp.(X.2).Q}
  \trans{\equiv}{Substituting value of $X.2$ and by $wlp$}
  \step{P \land \neg Q \land y ~\imp~ Q[pc.X \asgn 3]}
  \trans{\equiv}{By $wlp$ calculation}
  \step{P \land \neg Q \land y ~\imp~ true}
  \trans{\equiv}{By logic}
  \step{true}
\end{derivation}

\noindent
It is still required that $pc.X=2 \land \neg y \leadsto pc.X=3$ be
shown to complete the proof, which demands proof that a disabled
component makes progress. Transitivity may be used so that the proof
obligation is broken up into
\begin{eqnarray}
  \label{eq:1}
  pc.X=2 \land \neg y &\leadsto& pc.X=2 \land y  \\
  \label{eq:2}
  pc.X=2 \land y &\leadsto&  pc.X=3
\end{eqnarray}
so that becoming enabled, and making progress is shown in two
different steps. We will skip the proof of (\ref{eq:2}) as it is
similar to that of (\ref{eq:0}), however point out the importance of
our topology constraint in the proof of (\ref{eq:2}). Imagine that
there was a statement in component $Y$ that makes $y$ false. In our
proof of $pc.X = 2 \land y \un pc.X = 3$, we would have a calculation
of the following form:
\begin{derivation}
  \step{pc.X=2\land y \land pc.X \neq 3 ~\imp~ wlp.(i: y \asgn
    false~j:).((pc.X =2 \land y) \lor pc.X = 3)}
  \trans{\equiv}{By $wlp$ calculation and logic}
  \step{pc.X=2\land y ~\imp~ pc.X = 3}
\end{derivation}
which is clearly not true. 

We now return to the proof of (\ref{eq:1}).  Notice that component $X$
is disabled, which gives us no choice but to consult component $Y$.
Hence, we have:

\begin{derivation}
    \step{pc.X=2 \land \neg y ~\leadsto~ pc.X=2 \land y}
    \trans{\follows} {By disjunction}
    \step{(\all i ~:\!:~ pc.X=2 \land \neg y \land pc.Y=i ~\leadsto~
      pc.X=2 \land y)} 
\end{derivation}

\noindent
This is now demanding that the execution of the rest of the program,
i.e., component $Y$, lead to a state which makes the guard at $X.2$
true. We now perform case analysis on $pc.Y \in \{1,2,3\}$.
\begin{eqnarray}
  \label{eq:3}
  pc.X=2 \land \neg y \land pc.Y=1 & \leadsto& pc.X=2 \land y \\
  \label{eq:4}
  pc.X=2 \land \neg y \land pc.Y=2 &\leadsto& pc.X=2 \land y\\
  \label{eq:5}
  pc.X=2 \land \neg y \land pc.Y=3 &\leadsto& pc.X=2 \land y
\end{eqnarray}

\noindent
For (\ref{eq:3}), on account of $Y.1$ not hampering progress, on
account of being an assignment, and being orthogonal to
$pc.X=2 \land \neg y$, we opt for deferring the obligation to make
$y$ true, by delegating the task to $Y.2$. (\ref{eq:3}) is therefore
proved as follows:

\begin{derivation}
    \step{pc.X=2 \land \neg y \land pc.Y=1}
    \trans{\leadsto} {Immediate progress with $Y.1$}
    \step{pc.X=2 \land \neg y \land pc.Y=2}
    \trans{\leadsto} {By (\ref{eq:4})}
    \step{pc.X=2 \land y}
\end{derivation}

\noindent
For (\ref{eq:4}), since $Y.2$ is a guarded skip, deadlock is avoided by
requiring invariance of:
\begin{equation}
  \label{eq:I}
  pc.X=2 \land \neg y \land pc.Y=2 \land \neg x ~\imp~ false
\end{equation}

\noindent We may simplify (\ref{eq:I}) as follows: 
\begin{derivation}
    \step{pc.X=2 \land \neg y \land pc.Y=2 \land \neg x ~\imp~ false}
    \trans{\equiv} {By logic}
    \step{pc.Y=2 ~\imp~ pc.X \neq 2 \lor y \lor x}
\end{derivation}

\noindent
Since $pc.X$ is local to $X$ and because of the topological
constraint on $Y$, there is no choice but to introduce assignment
$y \asgn true$ at $Y.2$ to give us:

\cccode{.4} { The Initialisation Protocol -- refinement 1\\ 
  $Pre:~pc.X=pc.Y=1$ 
} { Component X: \\
  $
  \begin{array}[t]{ll}
    1: & y \asgn false;~ \\
    4: & x \asgn true;~ \\
    2: & \aang{\iif y \then~ skip~\Fi}~ \\
    3:
  \end{array}
  $
}
{ Component Y: \\
  $
  \begin{array}[t]{ll}
    1: & x \asgn false;~ \\
    4: & y \asgn true;~ \\
       & \{pc.X \neq 2 \lor y \lor x\}~ \\
    2: & \aang{\iif x \then~ skip~\Fi}~ \\
    3:
  \end{array}
  $
}

\noindent
(\ref{eq:4}) is now proved as follows:

\begin{derivation}
    \step{pc.X=2 \land \neg y \land pc.Y=2}
    \trans{\leadsto} {By (\ref{eq:I})}
    \step{pc.X=2 \land \neg y \land pc.Y=2 \land x}
    \trans{\leadsto} {Immediate progress rule with $Y.2$}
    \step{pc.X=2 \land \neg y \land pc.Y=3}
    \trans{\leadsto} {By (\ref{eq:5})}
    \step{pc.X=2 \land y}
\end{derivation}

\noindent
For (\ref{eq:5}), we opt to add a second assignment $y \asgn true$ at
$Y.3$ which gives us the following:

\cccode{.4} { The Initialisation Protocol -- refinement 2 \\
  $Pre:~ pc.X=pc.Y=1$ } 
{ Component X: \\
  $
  \begin{array}[t]{ll}
    1: & y \asgn false;~ \\
    4: & x \asgn true;~ \\
    2: & \aang{\iif y \then~ skip~\Fi};~ \\
    3: & x \asgn true~\\
    5:
  \end{array}
  $
}
{ Component Y: \\
  $
  \begin{array}[t]{ll}
    1: & x \asgn false;~ \\
    4: & y \asgn true;~ \\
    2: & \aang{\iif x \then~ skip~\Fi};~ \\
    3: & y \asgn true~\\
    5:
  \end{array}
  $
}

But note that this derivation is typical in its interplay between
proof and program development, and the new code at $Y.4$ and $Y.3$ has
extended the case analysis to cases $pc.Y \in \{1,2,3,4,5\}$. Case
$pc.Y=4$ is again by progress rule, but case $pc.Y=5$ is a different
matter. Evidently, introducing an assignment is not an option here for
reason of infinite regress, so we look to arrange invariance of:
\begin{equation}
  \label{eq:j}
pc.Y=5\imp y  
\end{equation}

\noindent
We now perform calculation on (\ref{eq:5}) which gives us: 

\begin{derivation}
    \step{pc.X=2 \land \neg y \land pc.Y=5}
    \trans{\leadsto} {By (\ref{eq:j})}
    \step{false}
    \trans{\leadsto} {Implication theorem}
    \step{pc.X=2 \land y}
\end{derivation}

\cccode{.4} { The Initialisation Protocol -- annotated for
progress
\\ $Pre:~ pc.X=pc.Y=1$} 
{ Component X: \\
  $
  \begin{array}[t]{ll}
    1: & y \asgn false;~ \\
    4: & x \asgn true;~ \\
    2: & \aang{\iif y \then~ skip~\Fi};~ \\
    3: & x \asgn true~ \\
    5: &
  \end{array}
  $
}
{ Component Y: \\
  $
  \begin{array}[t]{ll}
    1: & x \asgn false;~ \\
    4: & y \asgn true;~ \\
    2: & \aang{\iif x \then~ skip~\Fi};~ \\
    3: & y \asgn true~ \\
       & \{y\} \\
    5: &
  \end{array}
  $
}

\noindent
For GC of assertion $y$ at $Y.5$ we look to strengthen (\ref{eq:j}) with
$pc.X \neq 1$
\begin{equation}
  \label{eq:k}
  pc.Y=5 ~\imp~ y \land pc.X \neq 1
\end{equation}
\noindent
which induces the following annotation of $Y$:

\cccode{.4} { The Initialisation Protocol -- correctly annotated
\\ $Pre:~ pc.X=pc.Y=1$} { Component X: \\
  $
  \begin{array}[t]{ll}
    1: & y \asgn false;~ \\
    4: & x \asgn true;~ \\
    2: & \aang{\iif y \then~ skip~\Fi};~ \\
    3: & x \asgn true~ \\
    5: &
  \end{array}
  $
}
{ Component Y: \\
  $
  \begin{array}[t]{ll}
    1: & x \asgn false;~ \\
       & \{x \imp pc.X \neq 1\} \\
    4: & y \asgn true;~ \\
    2: & \aang{\iif x \then~ skip~\Fi};~ \\
       & \{pc.X \neq 1\} ~\\
    3: & y \asgn true~ \\
       & \{y\}\{pc.X \neq 1\}\\
    5:~
  \end{array}
  $
}

\noindent
GC of $pc.X \neq 1$ is for free because every action in $X$ makes
it true on account of $X.1$ being the initial action of $X$. This concludes the derivation. 

The example is a nice one for two reasons. First, because the problem
itself is quite delicate, as can be seen by reworking the design from
the point at which it was decided to establish (\ref{eq:3}) by the
transitivity rule rather than by introducing an assignment. The
alternative path leads all the way to:

\cccode{.4} { $Pre:~ pc.X=pc.Y=1$} { Component X: \\
  $
  \begin{array}[t]{ll}
    6: & x \asgn true;~ \\
    1: & y \asgn false;~ \\
    4: & x \asgn true;~ \\
    2: & \aang{\iif y \then~ skip~\Fi};~ \\
    3: & x \asgn true~ \\
    5: &
  \end{array}
  $
}
{ Component Y: \\
  $
  \begin{array}[t]{ll}
    6: & y \asgn true;~ \\
    1: & x \asgn false;~ \\
    4: & y \asgn true;~  \\
    2: & \aang{\iif x \then~ skip~\Fi}; ~\\
    3: & y \asgn true~ \\
       & \{y\}\{? ~pc.X \neq 1\} \\
    5:~
  \end{array}
  $
}

\noindent
but now the derivation falls down on account of the (lack of) GC
of $pc.X \neq 1$ at $Y.5$.

Second, while the derivation is marked by a complete absence of
operational thinking, yet it was completely driven by progress
concerns. This is just what we want to see in a problem like this
where progress is of the essence. In this regard, it is
instructive to compare it to the derivation in \cite{Feijen99} and
to note there the authors closing remark that ``we have to admit
that, no matter how crisp the final solution turned out to be, its
derivation seems to be driven by hope and a kind of
opportunism.''(p212). In our view, this is not true in the present
case, rather we see this derivation as a small step toward our
larger goal of developing a method of program derivation in which
progress requirements are given equal consideration with safety
requirements.

\subsection{The guard conjunction lemma}\label{GCL}

The guard conjunction lemma (\cite{Feijen99}, pp118-120) describes
a correct program transformation by justifying the replacement of
a guarded skip with a (coarse-grained) conjunctive guard $B \land
C$ by a pair of (fine-grained) guarded skips with guards $B$ and
$C$, when $B$ is GC in the component in which the guarded skip
occurs. The lemma states that the transformation preserves the
safety and progress properties of the original program, and it is
noteworthy that the proof of the latter part is outside of the
scope of the basic theory of Owicki and Gries (as presented in
Section \ref{CoreOG}). Thus, we are told by Feijen and van
Gasteren  that the basic theory ``is not suited for proving
[progress]. Fortunately, Dr. J. Hooman proved it for us. He did so
by considering the sets of all possible computations that can be
evoked by the original and by the new system, respectively, and
then showing that the two systems have the same properties as far
as deadlock and individual progress are concerned. The proof is
not for free and we are grateful to him for having designed it for
us.'' (\cite{Feijen99}, p118).  The purpose of this section is to
show how the guard conjunction lemma can be proved in the extended
theory of Owicki and Gries (as presented in Section \ref{ProOG}).
The lemma states

\begin{lemma*}[Guard Conjunction Lemma]
For a globally correct $B$, guarded command
\[i:\Angle{\iif B \land C \then S~ \Fi}~j:\]
may be replaced by
\[i:\Angle{\iif B \then skip~ \Fi};~ k:\Angle{\iif C \then S ~\Fi}~ j:\]
without impairing $total$ correctness of the design, i.e.
\begin{enumerate} 
\item [(i)] impairing the correctness of the annotation of the program
\item [(ii)] introducing total deadlock
\item [(iii)] endangering individual progress, i.e., given any
  component $X$, and labels $ii$, $jj$, each proof of the form
  $pc.X=ii \leadsto pc.X=jj$ is preserved
\end{enumerate}
\end{lemma*}
\noindent
For the sake of completeness, we begin by reproducing the proof of
(i). 

\proof We prove (i) as follows using notation $X \Refines Y$ to mean
``fragment $X$ can be transformed to fragment $Y$ without affecting
safety''.
\begin{derivation}
    \step{i: \Angle{\iif B \land C \then S~ \Fi} ~j: }
    \trans{\Refines} {Adding a $skip$ does not affect safety}
    \step{i: \Angle{\iif true \then skip~ \Fi};
          k: \Angle{\iif B \land C \then S~ \Fi}~j:}
    \trans{\Refines} {Strengthening the guard}
    \step{i: \Angle{\iif B \then skip~ \Fi};
          k: \Angle{\iif B \land C \then S~ \Fi}~j:}
    \trans{\Refines} {Introducing globally correct assertion $B$}
    \step{i: \Angle{\iif B \then skip~ \Fi}; \{B\}~
          k: \Angle{\iif B \land C \then S~ \Fi}~j:}
    \trans{\Refines} {Logic}
    \step{i: \Angle{\iif B \then skip~ \Fi}; \{B\}~
          k: \Angle{\iif C \then S~ \Fi}~j:}
    \trans{\Refines} {Weakening the annotation}
    \step{i: \Angle{\iif B \then skip~ \Fi};
          k: \Angle{\iif C \then S~ \Fi}~j:}
\end{derivation}
\noindent
Part (ii) follows from (iii) when we interpret (iii) to mean that a
program that contains the refined code has the $same$ progress
properties as the original program. In order to formalise (iii), we
conceptualise two programs, one is the original program that consists
of component $A$ and all other components. The other is this program,
but with $A$ replaced by $A'$, which is obtained from $A$ by replacing
the coarse-grained guarded skip by the pair of finer-grained
statements. 

We will first show that individual progress holds in the
modified component, then show that individual progress holds for
any other component in the program. Hence, we first show that: 
\[
(pc.A=i \leadsto pc.A=j) ~~~ \imp~~~ (pc.A'=i \leadsto pc.A'=j)
\]
Observe that when $B$ is globally correct, as the two codes
$\{B\}~\Angle{\iif C \then S~\Fi}$ and $\Angle{\iif B \land C \then S~
  \Fi}$ are equivalent, it is enough to prove that codes
\[i:\Angle{\iif B \land C \then S~ \Fi}~j:\]
and
\[i:\Angle{\iif B \then skip~ \Fi};~ k:\Angle{\iif B \land C \then S
  ~\Fi}~ j:\] 
have the same progress properties. Start by assuming that $A$ can pass
its guarded statement, i.e., $pc.A=i \leadsto pc.A=j$ holds. By the
immediate progress rule, this is only possible if the following 
equations hold:
\begin{center}
  $\begin{array}[t]{rll}
  pc.A=i & \leadsto &  pc.A=i \land B \land C \\
  pc.A=i \land B \land C & \leadsto &  pc.A=j
  \end{array}$
\end{center}

\noindent
We would thus like to show that the corresponding equations hold for
component $A'$ in the new program, i.e., prove that:
\begin{eqnarray}
  \label{eq:8}
  pc.A'=i & \leadsto &  pc.A'=i \land B \land C \\
  \label{eq:9}
  pc.A'=i \land B \land C & \leadsto&  pc.A'=j
\end{eqnarray}

In the new program, the only component that has changed is $A$, hence,
the rest of the program will preserve the proofs of (\ref{eq:8}) and
(\ref{eq:9}). Component $A'$ preserves the proof of (\ref{eq:8}) as
control remains at $i$, which completes the proof of (\ref{eq:8}).
Now, $A'$ is guaranteed to reach control point $k$ because:

\begin{derivation}
    \step{pc.A'=i \land B \land C}
    \trans{\leadsto} {Immediate progress rule, $B$ is GC in $A'$}
    \step{pc.A'=k \land B \land C}
\end{derivation}

\noindent
Furthermore, $pc.A'=i \leadsto pc.A'=k$ does not change the state of
the rest of the program, because a guard evaluation action can
only change the control state of the component in which the action
occurs, which is $A'$. Hence,

\fbox{
  \begin{tabular}[t]{lp{.8\columnwidth}}
    $CS:$ & When $pc.A=i$ and  $pc.A'=k$ the rest of the
              program containing $A$ is in the 
              same state as the
              rest of the program containing $A'$. 
  \end{tabular}
}

\noindent
which gives us the following calculation:

\begin{derivation}
    \step{pc.A'=k \land B \land C}
    \trans{\leadsto} {$CS$,~ $pc.A=i \land B \land C ~\leadsto~ pc.A=j$}
    \step{pc.A'=j}
\end{derivation}

\noindent
This concludes the proof that $A'$ is no less progressive than
$A$. We now prove that the implication also holds the other way, that is 
\[pc.A'=i \leadsto pc.A'=j ~~~ \imp~~~ pc.A=i \leadsto pc.A=j\]
\noindent
The above proof shows how $A'$ (with $pc.A'=k$) can get ahead of
$A$ (with $pc.A=i$) when $\neg C$ is true, but, of course, the
action at $A.k$ must wait for the rest of the program to make $C$
true. Since $pc.A'=k \land C \imp pc.A'=k \land B \land C$ by the
annotation of $A'$, and by $CS$, the $A.i$ guard is enabled
whenever the $A'.k$ guard is, which concludes the proof that $A'$
is no more progressive than $A$.

Now given any other component say $Y \neq A$, if the proof of $pc.Y=ii
\leadsto pc.Y=jj$ depends on the proof of $pc.A=i\leadsto pc.A=j$,
then as $pc.A'=i \leadsto pc.A'=j$ holds whenever $pc.A=i \leadsto
pc.A=j$, the assertion $pc.Y=ii \leadsto pc.Y =jj$ continues to hold
in the new program. If the proof does not depend on $A$, then as no
modifications have been made to the rest of the program, again $pc.Y =
ii \leadsto pc.Y = jj$ continues to hold.\qed

\section{Conclusion}\label{Conc}

In the context of sequential programs, Hoare \cite{Hoare69} showed how
a sequential program could be verified without reference to its
operational semantics. Then, in the context of concurrent programs,
Owicki and Gries \cite{Owicki76} showed how safety properties could be
verified by adding \emph{interference freedom} conditions to Hoare's
logic, but leaving the underlying logic unchanged. Although this
modification was small, the Owicki-Gries theory improved on the
previously existing global invariant method of \cite{Ashcroft75}
because it avoided a {\it state explosion} problem \cite{deRoever01}
by decomposing a global invariant into a program annotation
\cite{Lamport87}. In this paper we have developed this theory further
and incorporated a theory of progress into the formalism.

\cite{Owicki82} presents a proof system where the temporal operators
$\Box$ and $\Diamond$ have been incorporated into the Owicki-Gries
formalism. One of the drawbacks of their system is that both
conditional selection and blocking statements have not been described,
and one must simulate these using the looping construct.  The logic is
also missing both `next' and `unless' making it less expressive than
ours. Furthermore, keywords, `at', `after' and `in' are used to
describe the control state of the program, and temporal logic has been
encoded directly (as opposed to axiomatically) into their logic.
This has meant that the method needs to stay within the realms of
logical reasoning, as opposed to algebraic calculation, which as
\cite{Feijen99} has pointed out, is not suitable in the context of
program derivation.

Several event based models exist, such as
\cite{Chandy88,Back89,Lynch89,Shankar93,Lamport94}, but, as Lamport
suggests, proofs in these models can easily be translated from one
model to another, and the difference lies in the ease with which a
given program can be formalised in a given model. If a target
implementation is based on a concurrent sequential program model, then
we see no reason why this implementation should be modelled in an
event based one. We therefore see one advantage of our approach over
these others in the way that it can support a more direct translation
of a program design into code.

The extended theory of Owicki and Gries includes a logic of progress,
but it is up to us how to make use of it. Our ultimate aim is to
integrate this logic into a method of program design (derivation) in
the same kind of style as \cite{Feijen99}. Early work in this
direction is promising, and a more comprehensive example can be found
in \cite{Goldson05}, which presents a derivation of Dekker's program
for two process mutual exclusion. In a program verification, we do not
have the freedom to change a program when a proof does not work out.
We are left with the dilemma of not knowing whether the program or the
proof is at fault. In this respect, deriving a program that satisfies
a specification is certainly superior. \cite{Feijen99} have already
shown how commonly occurring design patterns can be identified in both
programs and their proofs, and how these patterns can be used to
shorten proofs. We believe that patterns such as these will emerge
with the extended theory as well. It is a case of realising when they
do and noting them accordingly.

We note that although $leads$-$to$ is a widely accepted construct for
reasoning about progress, it is not without deficiencies. For
instance, while $leads$-$to$ can always be used to express the
proposition that $P$ will {\em eventually} be true, by itself it can
not express the proposition that $P$ will be true in the {\em next}
program state. \cite{Shankar93} hints at the possibility of using
auxiliary variables to express the notion of {\em next} state. Whether
greater expressivity of temporal logic can be achieved in the
Owicki-Gries theory by combining auxiliary variables and $leads$-$to$
is a topic of further research.\\

\noindent
{\bf Acknowledgements}. We thank our anonymous referees for their
helpful comments and corrections. 

\bibliography{eog-lmcs}

\newpage

\section*{Appendix A: Derived rules of the logic of progress}

The logic of UNITY in \cite{Chandy88,Misra01} is based on an
inductive definition of a relation $leads$-$to$. Given this
definition, a number of derived properties are proved. The purpose
of this appendix is to confirm that these are also derived rules
of our progress logic too.  The Inductive Progress Rules in our
definition of $leads$-$to$ (in Section \ref{ProOG}) are identical
to those in \cite{Chandy88}, only the Immediate Progress Rule is
different, to take account of the different programming models.
Therefore, in what follows the proof of a derived rule will assume
that a use of $leads$-$to$ always results from a use of the Immediate
Progress Rule.

\begin{theorem*}[Implication Theorem]
  $(P \imp Q) ~\imp~ (P \leadsto Q)$
\end{theorem*}
  
\proof First note that, for any $R$, $(P \imp Q) ~\imp~ ((P \land
  \neg Q) \imp R)$. It follows that the three premisses of the
  Immediate Progress Rule are true on account of this equation,
  because 
\begin{enumerate}
\item
  \begin{derivation*}
    \step{P \un Q}
    \trans{\equiv}{By definition, for any atomic statement $\{U\}~S$}
    \step{P \land \neg Q \land U \imp wlp.S.(P \lor Q)}
  \end{derivation*}
\item $P \land \neg Q \imp pc = i$
\item Any atomic statement can be chosen on account of $P \land \neg Q
  \equiv false$ when assuming $P \imp Q$.\qed
\end{enumerate}

\begin{theorem*}[Impossibility Theorem]
$(P \leadsto false) \imp \neg P$
\end{theorem*}
\proof First note that, for any $S$, $wlp.S.false \equiv false$.
We look at the three forms of atomic statement that occur in
premiss (3) of the Immediate Progress Rule.

\begin{enumerate}
\item If $(P \leadsto false)$ because of a $skip$, assignment
  statement, or coarse-grained atomic statement $i:S~j:$ then
  \begin{derivation*}
    \step{P \imp wlp.S.false}
    \trans{\equiv}{By $wlp$ and logic}
    \step{\neg P}
  \end{derivation*}

\item If $(P \leadsto false)$ because of an $IF$ statement of the form

  $\quad i:\iif B_1 \then j:S_1 \oor B_2 \then k:S_2~\Fi~l:$ 
  \\then

  \begin{derivation*}
    \step{(P \land B_1 \imp false) ~\land~ (P \land B_2 \imp false)}
    \trans{\equiv}{By logic}
    \step{(\neg P \lor \neg B_1) ~\land~ (\neg P \lor \neg B_2)}
    \trans{\equiv}{By logic}
    \step{\neg P ~\lor~ (\neg B_1 \land \neg B_2)}
    \trans{\imp}{By premiss (3bi), $P \imp B_1 \lor B_2$}
    \step{\neg P}
  \end{derivation*}
\item If $(P \leadsto false)$ because of a $DO$ statement of the form
  
  $\quad i: \Do B \then j: S ~\Od~k:$ \\then
  \begin{derivation*}
    \step{(P \land B \imp false) ~\land~ (P \land \neg B \imp false)}
    \trans{\equiv}{By logic}
    \step{(\neg P \lor \neg B) ~\land~ (\neg P \lor B)}
    \trans{\equiv}{By logic}
    \step{\neg P ~\lor~ (\neg B \land B)}
    \trans{\equiv}{By logic}
    \step{\neg P\hbox to365 pt{\hfill\qEd}}
  \end{derivation*}
\end{enumerate}

\begin{theorem*}[Disjunction Theorem]
  \[(\all m~: m \in W :~ P.m \leadsto Q.m) ~\imp~
  ((\exists m ~: m \in W :~ P.m) \leadsto (\exists m ~: m \in W :~ Q.m))\]
\end{theorem*}
\proof
  As in \cite{Chandy88}.\qed

\begin{theorem*}[Cancellation Theorem]
\[(P \leadsto Q \lor D) \land  (D \leadsto R) ~\imp~ (P \leadsto Q \lor R)\]
\end{theorem*}
\proof
  As in \cite{Chandy88}.\qed

\begin{theorem*}[PSP Theorem]
  \[(P \leadsto Q) \land (R \un D) ~\imp~ (P \land R \leadsto (Q \land R)
  \lor D)\]
\end{theorem*}
\proof We assume the antecedent and show that, for the
  consequent, the three premisses of the Immediate Progress Rule are
  true. The proof uses two equations
  
  \begin{eqnarray}
    \label{eq:a1}
    R \land \neg Q \land \neg D & \equiv &  R \land
    \neg((Q \land R) \lor D)  \\
    \label{eq:a2}
    Q \land (R \lor D) & \imp &  (Q \land R) \lor D
  \end{eqnarray}
  
  \begin{enumerate}
  \item
    \begin{derivation*}
      \step{(P \leadsto Q) ~\land~ (R \un D)}
      \trans{\imp}{By Immediate Progress Rule}
      \step{(P \un Q) ~\land~ (R \un D)}
      \trans{\equiv}{By definition of $\un\!\!$, for any atomic
        statement $\{U\}~S$} 
      \step{(P \land \neg Q \land U \imp wlp.S.(P \lor Q)) ~\land~
        (R \land \neg D \land U \imp wlp.S.(R \lor D))}
      \trans{\imp}{By logic and conjunctivity of $wlp$}
      \step{P \land \neg Q \land R \land \neg D \land U ~\imp~
        wlp.S.((P\lor Q) \land (R \lor D))}
      \trans{\equiv}{By logic}
      \step{P \land \neg Q \land R \land \neg D \land U ~\imp~
        wlp.S.((P\land R) \lor (Q \land R) \lor (P \land D) \lor (Q \land D))}
      \trans{\imp}{As $(P \land D) \lor (Q \land D) ~\imp~ D$}
      \step{P \land \neg Q \land R \land \neg D \land U \imp
        wlp.S.((P\land R) \lor (Q \land R) \lor D)}
      \trans{\equiv}{By (\ref{eq:a1})}
      \step{P \land R \land \neg((Q \land R) \lor D) \land U ~\imp~
        wlp.S.((P\land R) \lor (Q \land R) \lor D)}
      \trans{\equiv}{By definition of $\un\!\!$}
      \step{P \land R ~\un~ (Q \land R) \lor D}
    \end{derivation*}
  \item For any statement $i:S~j:$
    \begin{derivation*}
      \step{(P \leadsto Q) ~\land~ R \un D }
      \trans{\imp}{By Immediate Progress and logic}
      \step{P \land \neg Q \imp pc = i}
      \trans{\imp}{By logic}
      \step{P \land R \land \neg ((Q \land R) \lor D) \imp pc = i}
    \end{derivation*}
  \item
    \begin{enumerate}
    \item If $(P \leadsto Q)$ because of a $skip$, assignment
      statement, or coarse-grained atomic statement $i: S~ j:$ then
      \begin{derivation*}
        \step{(P \leadsto Q) ~\land~ (R \un D)}
        \trans{\imp}{By definition of\un and $\leadsto$}
        \step{(P \land \neg Q \imp wlp.(i: S~ j:).Q) ~\land~
          (R \land \neg D \imp wlp.(i: S~ j:).(R\lor D))}
        \trans{\imp}{By logic}
        \step{P \land \neg Q \land R \land \neg D ~\imp~ wlp.(i: S~
          j:).(Q \land (R \lor D))} 
        \trans{\imp}{By (\ref{eq:a1}), (\ref{eq:a2}) and monotonicity
          of $wlp$}
        \step{P \land R \land \neg ((Q \land R) \lor D) ~\imp~ wlp.(i:
          S~ j:).((Q \land R) \lor D)} 
      \end{derivation*}
\newpage
    \item If $(P \leadsto Q)$ because of an $IF$ statement of the form 

      $\quad i: \iif B_1 \then j: S_1 \oor B_2 \then k: S_2~\Fi~l:$

      For premiss (3bi) 
      
      \begin{derivation*}
        \step{P \leadsto Q}
        \trans{\imp}{By definition}
        \step{P \land \neg Q ~\imp~ B_1 \lor B_2}
        \trans{\imp}{By logic}
        \step{P \land R \land \neg ((Q \land R) \lor D) ~\imp~ B_1 \lor B_2}
      \end{derivation*}
      
      and for premiss (3bii)
      
      \begin{derivation*}
        \step{(P \leadsto Q) ~\land~ (R \un D)}
        \trans{\imp}{By definition}
        \step{(R \land \neg D \land B_1 \imp (R \lor D)[pc \asgn j]) ~\land~
          (R \land \neg D \land B_2 \imp (R \lor D)[pc \asgn k]) ~\land~ }
        \step{(P \land \neg Q \land B_1 \imp  Q[pc \asgn j]) ~\land~
          (P \land \neg Q \land B_2 \imp  Q[pc \asgn k])}
        \trans{\imp}{By logic}
        \step{(P \land \neg Q \land R \land \neg D \land B_1
          \imp (Q \land (R \lor D))[pc \asgn j]) ~\land }
        \step{(P \land \neg Q \land R \land \neg D \land B_2
          \imp (Q \land (R \lor D))[pc \asgn k])}
        \trans{\imp}{By (\ref{eq:a1}) and (\ref{eq:a2})}
        \step{(P \land R \land \neg ((Q \land R) \lor D) \land B_1
          \imp ((Q \land R) \lor D)[pc \asgn j]) ~\land }
        \step{(P \land R \land \neg ((Q \land R) \lor D) \land B_2
          \imp ((Q \land R) \lor D)[pc \asgn k])}
      \end{derivation*}
      
    \item The case where $(P \leadsto Q)$ because of a $DO$ statement
      is similar to case (b).\qed
    \end{enumerate}
  \end{enumerate}

\begin{theorem*}[Induction Theorem] Let $M$ be a total function from
  program states to set $W$. Let $(W,<)$ be well-founded. Variable $m$
  in the following premiss ranges over $W$ and predicates $P$ and $Q$
  do not contain free occurrences of variable $m$.
  \[(\all m~:\!:~ P \land M=m \leadsto (P \land M<m) \lor Q) ~\imp~ (P
  \leadsto Q)\] 
\end{theorem*}
\proof
  As in \cite{Chandy88}.\qed

\begin{theorem*}[Completion Theorem] Let $P.i$ and $Q.i$ be predicates
  where $i$ ranges over a finite set.
  \[(\all i~:\!:~ (P.i \leadsto Q.i \lor D) \land (Q.i \un D))
  \imp~ ((\all i~:\!:~ P.i) \leadsto (\all i~:\!:~ Q.i) \lor D) \]
\end{theorem*}
\proof
  As in \cite{Chandy88}.\qed

\end{document}